\documentclass[conference, letterpaper]{IEEEtran}

\usepackage{amsmath,amssymb,amsfonts}
\usepackage{subfig}
\usepackage{algorithmic}
\usepackage{graphicx}
\usepackage{textcomp}
\usepackage{xcolor}
\usepackage{cite}
\usepackage{comment}
\usepackage{tabularx}
\usepackage{multirow}
\usepackage[euler]{textgreek}
\usepackage[keeplastbox]{flushend}
\usepackage[utf8]{inputenc}

\begin{document}

\title{\huge Event-Driven Source Traffic Prediction in Machine-Type Communications Using LSTM Networks}

	\author{\IEEEauthorblockN{Thulitha Senevirathna\IEEEauthorrefmark{1}\thanks{This Research work is funded by the Department of Electrical and Information Engineering, University of Ruhuna and the Telecommunications Regulatory Commission of Sri Lanka (TRCSL) and by the Academy of Finland 6Genesis Flagship under Grant 318927 and, in part.}, Bathiya Thennakoon\IEEEauthorrefmark{1}, Tharindu Sankalpa\IEEEauthorrefmark{1}, Chatura Seneviratne\IEEEauthorrefmark{1}, \\ Samad Ali\IEEEauthorrefmark{2}, and Nandana Rajatheva\IEEEauthorrefmark{2}} \\
		\IEEEauthorblockA{\IEEEauthorrefmark{1}
			Department of Electrical and Information Engineering, University of Ruhuna, Galle, Sri Lanka\\
			Email: thulithats@gmail.com, bathiyathennakoon@outlook.com, wltsankalpa@gmail.com, chatura@eie.ruh.ac.lk }
			
		\IEEEauthorblockA{\IEEEauthorrefmark{2}
			Center for Wireless Communications (CWC), University of Oulu, Oulu, Finland,\\
			Emails: \{samad.ali, nandana.rajatheva\}@oulu.fi}

	}	
		
\maketitle

\begin{abstract}
Source traffic prediction is one of the main challenges of enabling predictive resource allocation in machine-type communications (MTC). In this paper, a long short-term memory (LSTM) based deep learning approach is proposed for event-driven source traffic prediction. The source traffic prediction problem can be formulated as a sequence generation task where the main focus is predicting the transmission states of machine-type devices (MTDs) based on their past transmission data. This is done by restructuring the transmission data in a way that the LSTM network can identify the causal relationship between the devices. Knowledge of such a causal relationship can enable event-driven traffic prediction. The performance of the proposed approach is studied using data regarding events from MTDs with different ranges of entropy. Our model outperforms existing baseline solutions in saving resources and accuracy with a margin of around 9\%. Reduction in random access (RA) requests by our model is also analyzed to demonstrate the low amount of signaling required as a result of our proposed LSTM based source traffic prediction approach.   \end{abstract}

\begin{IEEEkeywords}
fast uplink grant, machine type communications, source traffic prediction, internet of things, long short-term Memory, recurrent neural network
\end{IEEEkeywords}

\section{Introduction}
Internet of Things (IoT) is expected to play a prominent role in the digital transformation of the societies. IoT will enable applications and services such as smart homes, smart grids, unmanned air crafts, and self-driving cars and improve the quality of life. To achieve this a large number of machine-type devices (MTDs) must be networked together, which will increase the traffic in communication networks. To handle a major part of this emerging traffic, the existing cellular networks seems to be a potential solution due to their already existing infrastructures, wide-area coverage, and high-performance capabilities \cite{TowardsaDatadrivenIoT}. However, the major obstacle for using the existing cellular networks for machine-type communications (MTC) is that the conventional random access (RA) procedure being designed to provide high data rates to devices in the downlink direction whereas, in MTC networks, short data packets are transmitted in networks with thousands of devices in the uplink direction \cite{TowardsaDatadrivenIoT}. Due to this reason, using the existing cellular networks in its current state for MTC would create traffic congestion and hence lead to collisions during transmissions \cite{ali2019fast}. 

There have been several methodologies proposed to overcome congestion at the cost of latency in prior art \cite{laya2013random,massivemtccom}. In contrast, fast uplink grant is a promising concept proposed by 3GPP \cite{3gpp.36.881} to overcome those limitations of conventional RA communication. This concept enables the base station (BS) to predict the set of transmission-ready MTDs beforehand to actively allocate uplink grant resources according to the Quality of Service (QoS) requirements\cite{ali2019fast}. For a successful prediction of source traffic, it is essential to consider two main types of traffic anticipated in an MTC environment, a) periodic and b) event-driven traffic. In periodic traffic, the MTDs tend to transmit at specific, predetermined time intervals. These time intervals could be ranging from several milliseconds to months. Calendar-based pattern mining is one of the ways to predict periodic and semi periodic traffic \cite{adhikari2013identifying}. Unlike periodic traffic, event-driven traffic does not have predetermined periods to initiate. The initiation is practically unpredictable. However, there are certain areas where there is a possibility of predicting traffic after the initiation. For instance, using proper tools, it is possible to predict the set of active MTDs that are casually activated after the first event-driven transmission. 

In \cite{ali2018directed}, authors use directed information (DI) to identify the causal connectivity between transmission patterns of different MTDs. The idea of directed information was initially used to find the capacity of a discrete channel with feedback \cite{massey1990causality}. However, its usefulness is proved in different applications including computational biology \cite{SeizureOnsetZone,GeneInfluence}. By using directed information, it is possible to infer the set of MTDs that are causally activated by the first transmission in an IoT event. Two series of transmissions from two MTDs are used in \cite{ali2018directed} to calculate a value of directed information for each time step in the series. So, the main focus behind this is to find the direction of causal connectivity between two devices and thereby get the direction where the event will propagate through time. However, in a single iteration of this calculation, the causality can be found only among two devices and in one direction. Thus, a higher number of devices would result in a large number of iterations for all possible permutations. Although directed information algorithm has the ability to identify event propagation direction, it cannot generate a sequence of predictions due to its nature.

The occurrence of IoT events in the network further contributes to these problems. Such an event that is not seen before is impossible to predict by looking at the transmission history. This is the main feature of event-driven traffic. Following the event, a burst of scheduling requests would occur in a legacy cellular network that uses RA scheduling requests only, causing congestion, delays, and wastage of resources\cite{ali2018directed}. However, under the concept of fast uplink grant, RA scheduling requests are only expected to be sent by the MTDs which are not identified by the base station beforehand as waiting-to-transmit devices. This reduces the signaling overhead and avoids most of the aforementioned problems. Our objective in this study is to predict MTDs waiting-to-transmit after an aforementioned event using an LSTM based approach.\par

The fundamental idea behind source traffic prediction is a simple causal relationship. However, when we take the context in application level, this problem could extend up to thousands of devices per base station. Identifying the causalities among all the devices using a rule/function based method could be computationally exhaustive in the long run due to the stochastic nature of their transmissions. Therefore, a statistical model such as a neural network would be more suitable to reduce the computational power needed. The intention behind this experiment is not merely to solve the problem of a few MTDs but also to stand as a starting point for scalable solutions for predicting source traffic. Another important reason behind selecting a machine learning model is to get better accuracy and precision. We have proved this from our results in section~\ref{secIV}. Artificial Neural Networks (ANNs) have proven capabilities in time series data analysis and recently been widely adopted in wireless communications problems \cite{ali20206g}. Our main focus in this paper is to use an LSTM network; a subset of ANN, to do source traffic prediction in fast uplink grant. With the rapid development of ANN technologies in the recent past, the prediction process of these networks demands a lower computation power. However, still the training of a neural network is computationally intense. With the advancement of infrastructure in cloud computers for deep learning, the training processes have also become much less of a burden. As a result, neural networks have become more suitable for time-series predictions videlicet the work at focus.\par

Our LSTM \cite{hochreiter1997long} network takes a restructured version of past transmission data and predicts future transmissions that experience similar IoT events. Artificial transmission data had to be generated to mimic MTC transmissions since real-world MTC network data was not publicly available at the time of our experiment. However, LSTM networks have a booming reputation for making predictions by identifying underlying correlations among input features. Therefore, the input data structure plays a paramount role in the final accuracy. Sequence lengths of input data are restructured beforehand to enable the model to identify important causalities among MTDs. Creation of the model took place in several stages to monitor the performance. Finally, hyper-parameters of the model were also fine-tuned to obtain maximum capacity with minimum computations. After a mild process of training, this model achieved higher accuracy in generating future sequences and lower RA resource wastage. The model presented here was also tested in a larger network without losing accuracy. The model's accuracy only tends to improve with the size of the data set, unlike in prior methodologies. So it is safe to mention that LSTM networks are more suitable for source traffic prediction in the fast uplink grant.

The rest of the paper is organized as follows. Section II presents the problem formulation. The proposed source traffic prediction model using LSTM is given in Section III. The performance evaluation is given in Section IV, and finally, conclusions are drawn in Section V. 

\section{Problem Formulation}
Consider the uplink of a cellular network with one base station (BS) and a set $\mathcal{M}$ of $M$ MTDs which are stationary or exhibiting low mobility. Transmission data is generated to mimic a scaled-down version of a simple MTC network. In a real-world MTC scenario, we assume that the BS has gathered a history of transmissions from MTDs using scheduling requests it received. Therefore, for each MTD, data is generated as a Bernoulli's distribution and it is arranged into time-series\cite{ali2018directed}. All the data is concatenated before restructuring data sequentially. After reshaping into adequate lengths of sequences, the data is reconstructed in a way to maximize comprehension of any causality among devices by the neural network. This plays a crucial part in the prediction of future transmissions. Thereafter data is split into three parts for three main phases of preparing a neural network: training, validation, and testing. However, in our case, we tune the model using a robust hyper-parameter tuning technique where training and validation tasks are simultaneously achieved. However, a separate testing phase is carried out to identify the prediction accuracy of the model.

\begin{figure}[]
\centering
\includegraphics[width=3.5in]{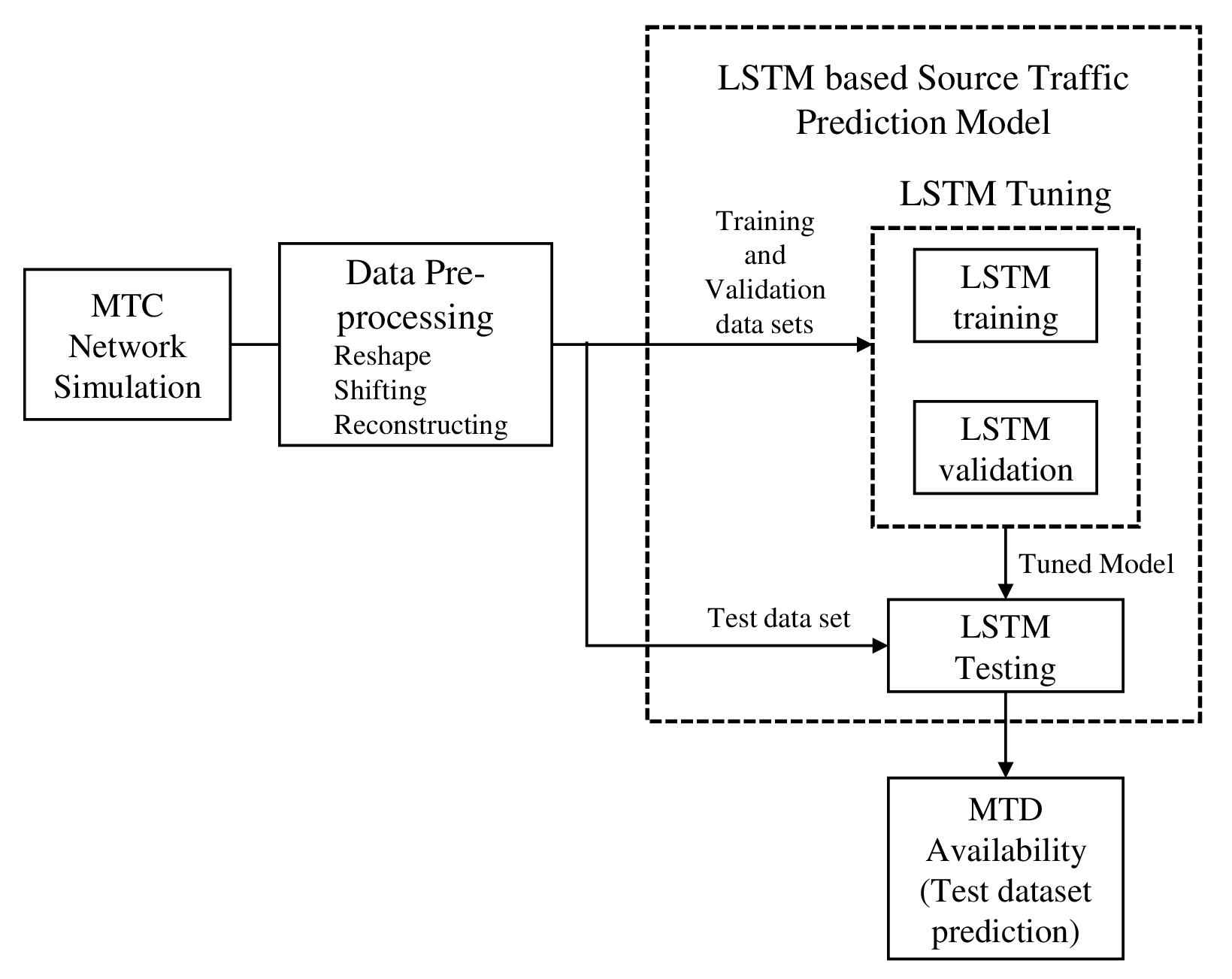}
\caption{Workflow from transmission data generation to prediction of future transmissions using the tuned LSTM model.}
\label{lstm_summary}
\end{figure}

During the testing phase, the performance of the model is evaluated using several standard performance metrics available for evaluating machine learning models. However, these metrics are used in a way to give a better insight into the model. The performance of the optimized model is compared against a conventionally trained model to convey the contrast. The workflow of the proposed LSTM based source traffic prediction model in the fast uplink grant of this cellular system is shown in Fig.~\ref{lstm_summary}.

\section{Source Traffic Prediction using LSTMs}
\subsection{Long Short Term Memory (LSTM) Model}

Out of all the machine learning approaches, we selected neural networks due to their ability to identify correlations among sequences in time series data sets. Recurrent neural networks (RNNs) are widely used as an effective model for identifying patterns and forecasting sequential data in many applications\cite{greff2016lstm}. In traditional RNNs, the gradient signal can end up being multiplied for a large number of times by the weight matrix associated with the connections causing vanishing/exploding gradient problems\cite{liu2018stock}. LSTMs have overcome these issues using a memory cell; which can maintain its state over time, and nonlinear gating units; which regulate the information flow into and out of the cell\cite{greff2016lstm}. \par

\begin{figure}
\centering
\includegraphics[trim = 0mm 0mm 0mm 0mm, clip, width=3.5in ,height=2.8in]{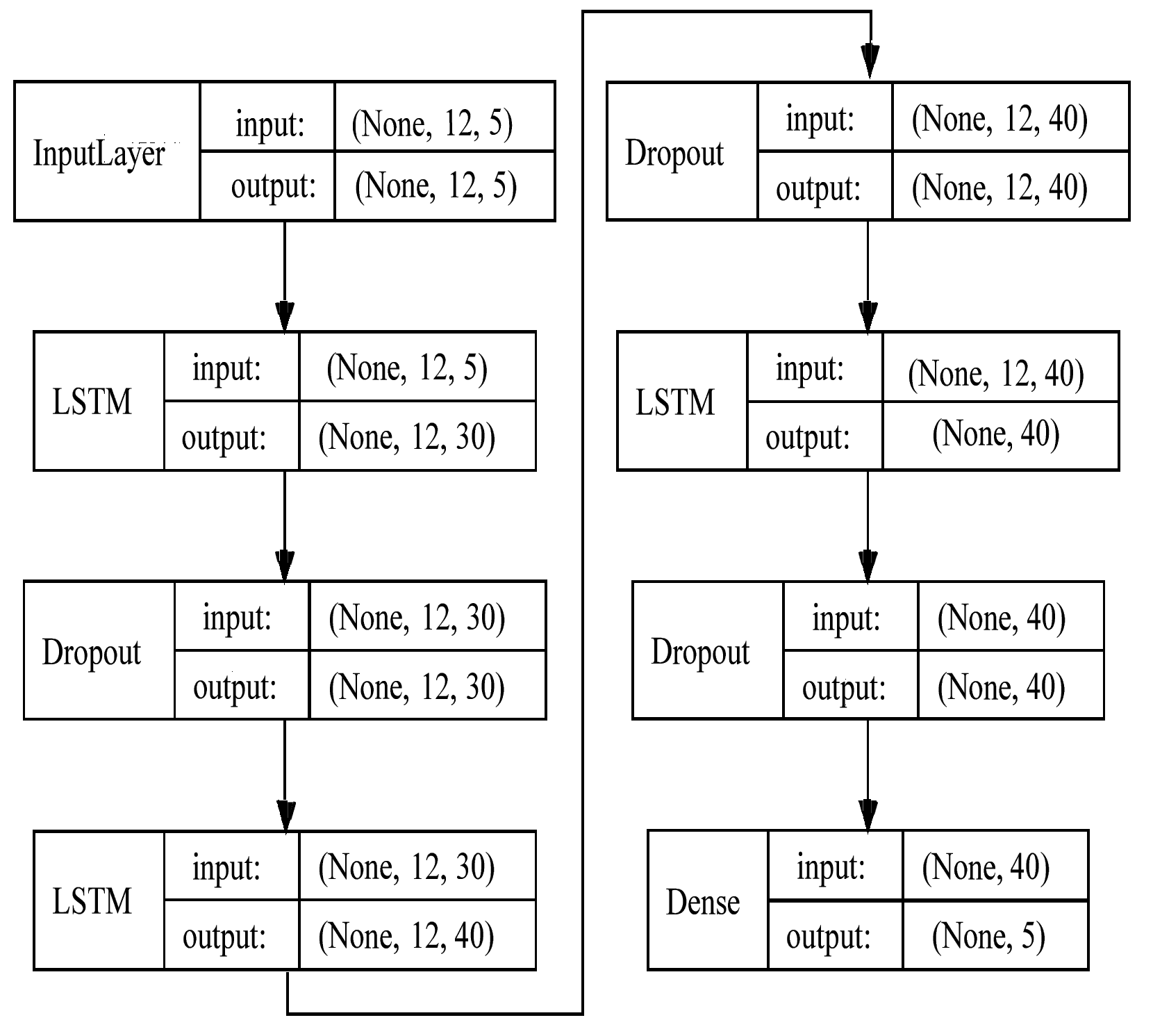}
\caption{LSTM model structure for 5 MTDs after tuning Bayesian hyper parameter optimization process.}
\label{LSTM_Structure}
\end{figure}

As shown in Fig.~\ref{LSTM_Structure}, our optimal LSTM model accepts a 3D array of sequences of length 12 for 5 devices where the number of sequences can be a variable. First, the LSTM layer outputs hidden states for each input time step and this continues until the final LSTM layer. However, the passing of hidden states is done through a dropout regularization layer where 27\% of the hidden states are dropped before sending the values. In the last LSTM layer, after dropout regularization, it only outputs one hidden state per sequence. Finally, a dense layer is used to compare labels with predictions.

The choice of loss function for this model is made considering the result of a hyper-parameter tuning job. However, the candidates for the loss function are selected based on well known conventions. Time stepwise, this might seems like a binary classification. Still it cannot be considered as such because of its undeniable regressive nature. Therefore, categorical loss functions such as log-loss and hinge-loss would render incongruous with this problem. We select Mean Square Error (MSE) to test the performance of the model without discarding the effect from potential outliers. We also use log-cosh function which is rather insensitive to sudden changes, to capture the contrast between the two approaches. Equations \ref{RMSE} and \ref{logcosh} represent MSE and Log-Cosh functions respectively,

\begin{equation}
    MSE =\frac{\sum_{i=1}^{n} (Y_i- \hat{Y}_i)^{2}}{n},
\label{RMSE}
\end{equation}

\begin{equation}
    L(Y,\hat{Y}) =\sum_{i=1}^{n} \log(\cosh(\hat{Y}_i-Y_i))
\label{logcosh}
\end{equation}

where $n$ is the number of training samples, $Y_i$ is the actual value, $\hat{Y}_i$ is the value predicted by the algorithm and $L(Y, \hat{Y})$ is the loss. From the sample of training jobs MSE shows promising results and thus it is used in the final model.

For learning rate optimization, we use Adam optimizer algorithm\cite{kingma2014adam} since it can leverage the power of adaptive learning rate methods by finding individual learning rates for each parameter. We use binary accuracy as the model training metric since our interest is in predicting a Bernoulli's distribution with our regression model.

\subsection{Transmission Data Generation}
We generate artificial transmission data to simulate the event-driven transmission patterns of the MTDs. Five MTDs were considered (X, Y, Z, T, W) for the data generation. For simulation purposes, the length of the sequence of an event is assumed to be 12 time steps.\par 

For MTD X, transmission happens at time steps t \(\in\) \{3, 4, 5, 6, 9, 10\}. In other words, at time steps 3, 4, 5, 6, 9, and 10, the MTD may or may not transmit. The probability of X to transmit at any of the given time steps is assumed to be constant in this case. Nevertheless, during the remaining time steps, it is assumed that the MTD will be in a state of sleep where no transmissions occur. An example transmission pattern of MTD X can be given as X\textsuperscript{12} = \{001101001000\}. For MTD Y, the random transmission happens at any time. In other words, there can be a transmission at any given time step during the 12 step sequence. Y causes T and W to transmit directly and indirectly respectively. All together, Z, T, and W exhibit an event-driven transmission nature correlated to X and Y. For Z, we assume that the transmission occurs 3 time steps after X with a probability of 0.7. For MTD T, the transmission happens 2 time steps after Y with a probability of 0.7. Finally, for W, we assume that the transmission occurs one time step after T with a probability of one. By adopting such a structure we can cover three levels of entropy; completely random (Y), partially uncertain (Z and T), and completely certain (W). Also in W, we can see a 2\textsuperscript{nd} degree causality with Y.\par

\subsection{Data Preprocessing}

\begin{figure}
\centering
\includegraphics[trim = 0mm 0mm 0mm 0mm, clip, width=3.5in]{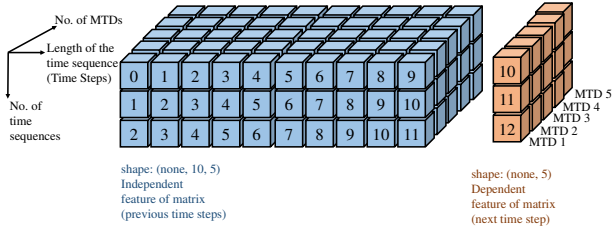}
\caption{An example structure of the input data set to LSTM model. Here the light colored time steps are input data while the shaded parts are labels.}
\label{input_data}
\end{figure}

Before feeding data to the LSTM, the generated data is reshaped optimally for the model to understand. Generally, the input of an LSTM takes the shape of a 3D array. The X, Y, and Z-axes of the array represents the number of sequences, the length of a sequence, and the number of MTDs respectively. This is clearly illustrated in Fig.~\ref{input_data}. All the event sequences from an MTD are reshaped into a vector and then they are arranged in a 2D matrix where the width is decided by the number of time steps per sequence. The longer vector generated initially represents a time series of transmissions arriving at a base station from an MTD. This is then broken down in to feature sequences that can be learned by the LSTM model. Breaking down into sequences is done in a sliding window approach to augment the data set and uncover more hidden patterns which could potentially be learned by the model alongside with causality among MTDs. Parameter for the optimal length of those sequences is found by fine-tuning it in the hyper-parameter tuning process. As an example structure, Fig.~\ref{input_data} shows a modified sequence length of 10. \par
The next single time step after the first 10, is taken as the label in the generated transmission data. They are given as 10\textsuperscript{th}, 11\textsuperscript{th}, 12\textsuperscript{th} and so on for MTD 1. During the training process the transmission pattern of previous 10 time steps are fed to the LSTM as the independent feature matrix alongside with the labels. It was then tuned later on to yield the best performing model.

\subsection{Model Training}
\begin{figure}
\centering
\includegraphics[trim = 0mm 0mm 0mm 0mm, clip, width=3.5in]{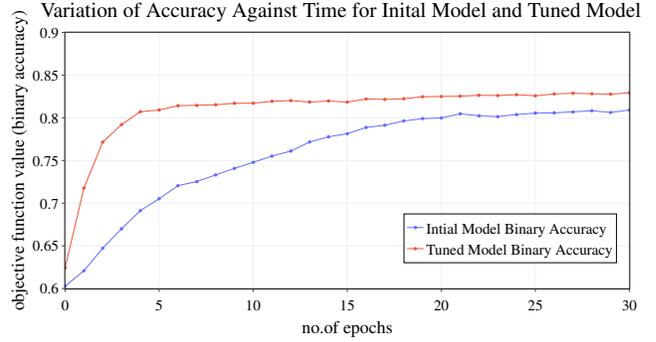}
\caption{Comparison between initial model and tuned model with regard to binary accuracy objective function.}
\label{Initial_model_accuracy}
\end{figure}

Training of the model is done in two main ways. The model is first trained with arbitrary parameters and then it is trained using a Bayesian Optimizer.
Fig.~\ref{Initial_model_accuracy} shows the accuracy variation of the very first training process. The layer sizes for the first and last layers are obtained based on the guidelines from literature \cite{huang2003learning}. The rest of the parameters are added arbitrarily or according to the general convention. This model is used for performance comparison to identify the difference between any improvements obtained with hyper-parameter tuning. Training of this model is not computationally demanding. Thus, any kind of acceleration is not needed. After the tuning process, the best model is trained with the tuning results. A comparison of convergence in binary accuracy from both models is given in Fig.~\ref{Initial_model_accuracy}. The tuned model is faster during the training process and converges to a higher accuracy level at the end of a lesser number of epochs than the initial model.

\section{Improvements Gained using LSTM Neural Network}\label{secIV}
\subsection{Model Tuning}
 The Bayesian optimizer is used for hyper-parameter tuning and validation purposes. It has a high reputation on finding the optimal set of hyper-parameters for any model from a given range of values more efficiently\cite{mishra_2019} than other traditional methods such as manual method, grid search, and random search. Internally, Bayesian hyper-parameter optimizer builds a probability model of the objective function and uses it to select the most promising hyper-parameters to evaluate the true objective function. By setting binary accuracy as the objective function, we have given pragmatic ranges for hyper-parameters of the model.\par
 Model tuning is done with two main intentions. First to fine-tune the model hyper-parameters and obtain the best binary accuracy. The results obtained under this tuning is given in Table.~\ref{hyperparameter_tuned} with ranges of hyper-parameters used. Secondly to obtain the optimal length of restructured sequence length. 
 
\begin{table}
\caption{Hyper-parameter Ranges used and Best Model Parameters}
\begin{center}
\begin{tabular}{>{\raggedright}p{3cm}>{
\raggedright}p{2cm}>{
\raggedright\arraybackslash}p{2cm}}
\hline
\textbf{Hyper parameter} &\textbf{Range} &\textbf{Best Model Parameters} \\
\hline
Batch Sizes& 32-128 & 113 \\
Dropout & 0.1-0.9 & 0.27 \\
Number of Epochs & 1-50 & 26 \\
Number of Hidden Layers & 1-5 & 3 \\
Number of Hidden Layer Units & 3-200 & 40 \\
Number of Input Hidden Layer Units & 10-200 & 30\\
Learning rate & 0.001 – 0.1 & 0.007\\
Loss Type &	Mean Squared Error,
Logcosh & Mean Squared Error \\
Optimizer& Adam,
RMSprop & Adam\\
Number of Dense Layer Units & - &5 \\
Objective metric & - & Binary Accuracy \\
\hline
\end{tabular}
\label{hyperparameter_tuned}
\end{center}
\end{table}

\begin{figure}[h]
\centering
\includegraphics[trim = 0mm 0mm 0mm 0mm, clip, width=3.5in]{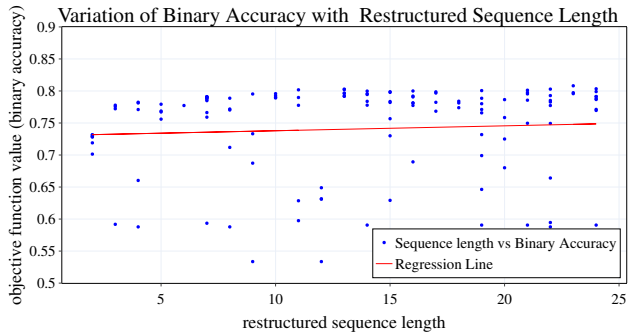}
\caption{Variation of Binary Accuracy Against Sequence Length. Effect from the sequence length can be compensated by other hyper parameters.}
\label{SequenceLengthVsBinAcc}
\end{figure}

\begin{table}[]
\caption{Performance of LSTM model for each MTD with different transmission patterns.}
\label{prediction_results}
\begin{tabular}{cllll}
\hline
\multicolumn{1}{l}{\textbf{MTD name}} & \textbf{Sensitivity} & \textbf{FDR} & \textbf{Accuracy} & \textbf{MCC} \\ \hline
MTD X & 82.50\% & 15.89\% & 83.38\% & 0.667652 \\
MTD Y & 59.60\% & 49.25\% & 50.92\% & 0.018856 \\
MTD Z & 99.76\% & 31.66\% & 84.09\% & 0.718287 \\
MTD T & 100.00\% & 32.05\% & 84.00\% & 0.717656 \\
MTD W & 100.00\% & 0.00\% & 100.00\% & 1.000000 \\ \hline
\end{tabular}
\end{table}

\subsection{Restructuring of Sequence Length}
In Fig.~\ref{SequenceLengthVsBinAcc}, the Bayesian optimizer has successfully identified configurations where the model could achieve an accuracy of around 80\% for almost all the lengths. From the graph we can see a slight increment in accuracy when the sequence length is increased. This is mainly due to the additional information stored in LSTM cells contributed by the longer sequences of data. However, for the sake of pragmatism, it is safe to say that despite the length of each restructured sequence, there is a possibility of achieving the best prediction accuracy even by adjusting the rest of the parameters properly. These parameters can be selected based on factors such as computational power available and other memory constraints. \par
Ultimately the model learns patterns of transmission from each MTD and underlying causalities of the MTDs that are transmitting in an event-driven situation. This was clear from the prediction results in Table~\ref{prediction_results}. Here it's apparent that even though the model performed poorly for MTD Y; a randomly transmitting device representing initiations of events, MTD T and W; which are causally dependent on Y, were predicted with a higher accuracy. Also with the prediction results from MTD X; a periodically transmitting MTD, LSTMs have proved that periodicity of transmissions by each MTD was also learnt in the process. As mentioned in \cite{ali2019fast}, the need for implementing various calendar-based techniques as a separate system for periodic traffic prediction would be redundant when LSTM neural networks are used.

\begin{figure}[]
\centering
\includegraphics[trim = 0mm 0mm 0mm 0mm, clip, width=3.16in]{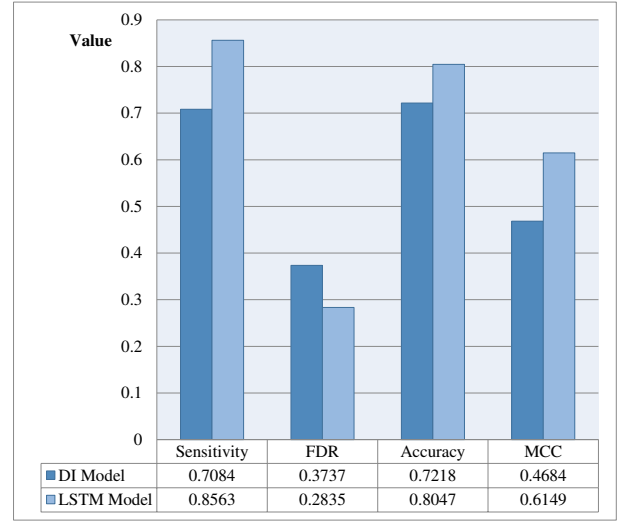}
\caption{Comparison of DI model vs LSTM model on performance metrics related to resource wastage and reduction of RA requests.}
\label{comparison_DI_LSTM}
\end{figure}

\subsection{Contribution towards minimizing resource wastage and RA requests.}

At the time of writing this paper, the DI system is the only publicly available solution for event-driven source traffic prediction. From Fig.~\ref{comparison_DI_LSTM} it is clear that under all performance metrics, the LSTM model outperforms the DI model. Interpretation of each metric can be done under four main criteria as shown in Fig.~\ref{comparison_DI_LSTM}. Mathew's correlation coefficient (MCC) and accuracy are arguably a very comprehensive measure of the prediction capability of any statistical model. From the 14\% higher MCC score, we can say that the LSTM model assigns values in a lesser random nature than the DI model. In other words, the LSTM model has a better quality prediction than DI, which is closer to being perfect. Accuracy, in its usual meaning, is 8\% higher than the DI model. \par
On the contrary, sensitivity has a more contextual meaning in this application. That is, it gives a clear fraction of how many transmissions are correctly predicted out of all actual transmissions. Every correct transmission in this context amounts to a reduction in RA requests, otherwise required to be sent. So effectively, a 15\% higher sensitivity means that the LSTM model has the potential to reduce the number of RA requests that have to be sent to the network.\par
Similarly, False Discovery Rate (FDR) also has a more contextual meaning. This represents the fraction of falsely predicted transmissions, out of all predicted transmissions. In other words, this value represents the potential number of allocations of resource blocks to sleeping MTDs. Therefore, these resource blocks are wasted, without being used by an MTD waiting for transmission. Thus, it is safe to say that FDR represents an indirect proportionality with the quality of the model. So, a 9\% decrease in resource wastage than DI can be considered as an improvement.

\section{Conclusion}
In this paper, we have presented a novel LSTM based approach to predict the event-driven source traffic in MTC networks. First, we have modeled the transmission history of the MTDs by a Binary time series. Then, we have proposed an LSTM based deep learning approach for inferring causal relations between the transmission patterns of the MTDs and predicting event-driven traffic. The results show that the LSTM model is capable of identifying the causalities between devices and outperforms the accuracy, extensibility, and scalability of the existing DI based prediction method. Therefore, the proposed LSTM based neural network is a promising approach for source traffic predictions to enable predictive resource allocation mechanisms for MTC in future wireless networks.

\section{Acknowledgment}
This Research work is in part funded by the Department of Electrical and Information Engineering, University of Ruhuna and the Telecommunications Regulatory Commission of Sri Lanka (TRCSL). This work was also supported in part by the Academy of Finland 6Genesis Flagship under grant 318927.

\newpage
\bibliographystyle{IEEEtran}
\bibliography{bibliography.bib}

\end{document}